\documentclass[prb,aps,showpacs,citeautoscript, superscriptaddress,amsmath,amssymb,floatfix,twocolumn,dvipsnames]{revtex4-1}
\usepackage{graphicx}
\usepackage{xcolor}
\usepackage{ytableau}
\usepackage[colorlinks,bookmarks=false,citecolor=blue,linkcolor=red,urlcolor=black]{hyperref}
\usepackage{amsfonts}
\usepackage{amsmath}
\usepackage{times}
\usepackage{amssymb}
\usepackage{changes}
\usepackage{blkarray}

\ytableausetup{smalltableaux} 

\setcitestyle{super}
\newcommand*{\citen}[1]{%
  \begingroup
    \romannumeral-`\x 
    \setcitestyle{numbers}%
    \cite{#1}%
  \endgroup   
}

\newcommand{\bra}[1]{\langle #1|}
\newcommand{\ket}[1]{|#1\rangle}

\usepackage{color}
\definecolor{lightpink}{RGB}{255, 187, 218}
\definecolor{lightblue}{RGB}{130, 215, 255}
\definecolor{lightgrey}{RGB}{220, 220, 220}
\definecolor{lightred}{RGB}{255, 155, 155}
\definecolor{lightteal}{RGB}{143, 255, 243}
\definecolor{lightindigo}{RGB}{191, 181, 255}
\definecolor{lightgreen}{RGB}{185, 255, 179}
\definecolor{lightpurple}{RGB}{237, 179, 255}

\definecolor{c1}{RGB}{144,12,63}
\definecolor{c2}{RGB}{199,0,57}
\definecolor{c3}{RGB}{255,87,51}
\definecolor{c4}{RGB}{255,195,0}
\definecolor{c5}{RGB}{218,247,166}

\begin{document}

\title{Exact diagonalization of $\mathrm{SU}(N)$ Heisenberg and AKLT chains using the full $\mathrm{SU}(N)$ symmetry}
\date{\today} 

\author{Kianna Wan}
\affiliation{Institute of Theoretical Physics, \'Ecole Polytechnique F\'ed\'erale de Lausanne (EPFL), CH-1015 Lausanne, Switzerland}
\affiliation{Department of Physics and Astronomy, University of Waterloo, Waterloo, Ontario N2L 3G1, Canada}
\author{Pierre Nataf}
\affiliation{Institute of Theoretical Physics, \'Ecole Polytechnique F\'ed\'erale de Lausanne (EPFL), CH-1015 Lausanne, Switzerland}
\author{Fr\'ed\'eric Mila}
\affiliation{Institute of Theoretical Physics, \'Ecole Polytechnique F\'ed\'erale de Lausanne (EPFL), CH-1015 Lausanne, Switzerland}

\begin{abstract}
We present a method for the exact diagonalization of the $\mathrm{SU}(N)$ Heisenberg interaction Hamiltonian, using Young tableaux to work directly in each irreducible representation of the global $\mathrm{SU}(N)$ group. This generalized scheme is applicable to chains consisting of several particles per site, with any $\mathrm{SU}(N)$ symmetry at each site. Extending some of the key results of substitutional analysis, we demonstrate how basis states can be efficiently constructed for the relevant $\mathrm{SU}(N)$ subsector, which, especially with increasing values of $N$ or numbers of sites, has a much smaller dimension than the full Hilbert space. This allows us to analyze systems of larger sizes than can be handled by existing techniques.
We apply this method to investigate the presence of edge states in $\mathrm{SU}(N)$ Heisenberg and AKLT Hamiltonians.

\end{abstract}

\maketitle

\section{I. Introduction}
In recent years, considerable progress has been made in experiments with ultracold atoms \cite{dalibard}, enabling the realization of sophisticated quantum many-body systems. In particular, degenerate gases of strontium and ytterbium loaded in optical lattices
have been used to simulate the $\mathrm{SU}(N)$ Fermi-Hubbard model \cite{WuPRL2003,Honerkamp2004,Cazalilla2009,gorshkov2010,takahashi2012,Pagano2014,Scazza2014,Zhang2014}, a generalization of the familiar $\mathrm{SU}(2)$  spin-$1/2$ Fermi-Hubbard model. When the number of particles per site is an integer and the on-site repulsion is sufficiently large, the systems are expected to be in Mott insulating phases, which are well-described by $\mathrm{SU}(N)$ Heisenberg models. This class of models is a unique playground for strongly correlated systems as it encompasses a wide variety of quantum ground states with different physical properties. In fact, even for the simplest cases with interactions limited to nearest neighbours, the zero-temperature quantum phases can be very diverse and can depend on the geometry of the lattices (one-dimensional chain, two-dimensional bipartite or frustrated lattice), the number of \emph{colours} (i.e., the value of $N$), and the local $\mathrm{SU}(N)$ symmetry of the wave function. 

At each site, the local $\mathrm{SU}(N)$ symmetry corresponds to a specific irreducible representation (``irrep") of $\mathrm{SU}(N)$, and, for $m$ particles per site, can be encoded by Young diagram with $m$ boxes and no more than $N$ rows. For $m=1$, the Young diagram is a single box, representing the fundamental irrep. In this case, the $\mathrm{SU}(N)$ chain, for which a general Bethe ansatz solution exists \cite{sutherland}, is gapless, with algebraic decaying correlations. However, if a second particle is added to each site in such a way that the resultant local wave function is fully symmetric, then, in the case of $N=2$, the system can open a Haldane gap\cite{haldanegap}, while for $N>2$, the chain should be critical with universality class  $\mathrm{SU}(N)_1$ (
although this issue has not yet been completely solved from a numerical point of view\cite{PRBNataf2016}). In two dimensions, the ground state of a square lattice with $m=1$ particle per site has been shown to be characterized by some N\'eel-type ordering for $\mathrm{SU}(2)$, $\mathrm{SU}(3)$\cite{toth2010,bauer_three-sublattice_2012}, $\mathrm{SU}(4)$\cite{corbozSU42011} and $\mathrm{SU}(5)$\cite{nataf2014}, whereas when there are $m>1$ particles per site in an antisymmetric representation, the ground state is predicted by mean-field theory to be a chiral spin liquid, provided that $m/N>5$ \cite{hermele2009,hermele_topological_2011}.

From an experimental perspective, the study of $\mathrm{SU}(N)$ Heisenberg models with $N>2$ and non-fundamental irreps (i.e., $m>1$) on each site is indeed relevant. It has been shown, for instance, that certain many-body systems exhibit $\mathrm{SU}(N)$ symmetry with $N$ as large as 10, as in the case of strontium-87\cite{Zhang2014,Scazza2014}, and may be implemented using optical lattices with two atoms on each site \cite{gorshkov2010,Cazalilla2014}. Furthermore, the realization of exotic phases of matter is naturally expected to involve irreps of mixed symmetry, that is, neither fully symmetric (corresponding to Young diagrams with one row) nor fully antisymmetric (corresponding to Young diagrams of one column). According to mean-field caculations\cite{hermele2009,hermele_topological_2011}, in order to obtain non-Abelian chiral spin liquids on the square lattice, the number of columns of the local irreps should be at least two. Another example is the $\mathrm{SU}(N)$ symmetry-protected topological phases in one dimension\cite{nonne2013,Capponi_annals_2016}, which are generalizations of the $\mathrm{SU}(2)$ spin-1 Haldane phases\cite{rachel2007,rachel2009,nonne2011}.
These are gapped phases with 
non-trivial
edge states, and the paradigmatic Hamiltonians of those states are the $\mathrm{SU}(N)$ version of the AKLT chain\cite{AKLT1,AKLT2}.%
They involve irreps with multiple rows and columns at each site and can lead to $N$ distinct topological phases, classified using group cohomology\cite{Quella2013_phases}. 

The theoretical study of such systems can be extremely challenging, due in no small part to the inherent limitations of current numerical methods. The Density Matrix Renormalization Group (DMRG) technique has proven rather efficient in the investigation of $\mathrm{SU}(N)$ Hamiltonians in 1D \cite{rachel2009,nonne2013,quella2012,fuhringer2008,manmana2011} as well as Infinite Projected Entangled Pair States (iPEPS) in 2D \cite{bauer2012,corbozPRX2012,corbozsimplex2012,corbozSU42011}. However, as the local Hilbert space dimensions increases as a result of increasing the number of colours or the number of particles per site, the performance of DRMG significantly deteriorates. Quantum Monte Carlo methods, on the other hand, are usually able to accommodate large Hilbert spaces, but can only be used in very specific cases (to avoid the sign problem), namely, for chains with one particle per site, or for bipartite lattices on which pairs of interacting sites correspond to conjugate irreducible representations\cite{capponiQMC,assaad2005,cai2013,lang2013,zhou2014}. For configurations where the wave function on each site is completely antisymmetric, variational Monte Carlo simulations based on Gutzwiller projected wave functions have been found to produce remarkably accurate results \cite{wang_z2_2009,paramekanti_2007,lajko_tetramerization_2013,dufourPRB2015,Dufour2016}, but it is not clear how this approach can be generalized to other local irreps. Finally, exact diagonalization (ED) is limited to small clusters.

Recently, we developed a procedure to exactly diagonalize the Hamiltonian for one particle per site independently in each global irrep of $\mathrm{SU}(N)$, using standard Young tableaux (SYTs) and the orthogonal representation of the symmetric group\cite{nataf2014}. This method obviates the use of Clebsch-Gordan coefficients, for which the computation complexity increases dramatically with $N$ \cite{Alex2011}. The method was then extended to chains with a fully symmetric or antisymmetric irrep at each site\cite{PRBNataf2016}---a rather straightforward extension, as the number of multiplets per site is still one. However, the number of multiplets is greater than one for systems where local irreps have more than one row and more than one column, and it was not clear how SYTs could be used to solve such systems.

\bigskip

The purpose of this article is to proceed to such a development, and the method presented in the following two sections can be applied to the most general configuration, with one or more particles per site and \emph{any} given irrep at each site (not necessarily the same from site to site). Following a brief introduction on the theory behind the method, in Section \ref{S: method} we show how to derive a \textit{projection operator} which imposes the local $\mathrm{SU}(N)$ symmetry at each site. Then, in Section \ref{S: algorithm}, we describe an efficient algorithm for constructing suitable basis states for a for a given global $\mathrm{SU}(N)$ subsector, using this projection operator. Further simplifications to this algorithm, which take full advantage of the inherent symmetry of the problem, are detailed in the Appendix. In Section \ref{S: results}, the method is used to investigate the 
$\mathrm{SU}(4)$ Heisenberg chain with irrep $[2,2]$ at each site: After demonstrating how to systematically express an $\mathrm{SU}(N)$ AKLT Hamiltonian in terms of permutations,
we calculate the energy of the edge states directly in their corresponding irreps and determine whether they remain in the lowest part of the spectrum as we move from the AKLT point to the Heisenberg point in an interpolating Hamiltonian. 
Finally, conclusions are drawn and future directions are discussed in Section \ref{S: conclusions}.

\bigskip

\section{The Method} \label{S: method}

\subsection{The Hamiltonian as a sum of permutations}

As in Ref. \onlinecite{nataf2014}, the construction of the Hamiltonian matrix will rely on the very simple representation of permutations in the basis of standard Young tableaux. It is therefore most convenient to define the model directly in terms of permutations. The equivalence of this to other formulations will be established in the following subsection.

Consider a general $\mathrm{SU}(N)$ model
where, at each site, there are $m$ particles in the fundamental representation. Denoting the number of sites by $N_s$, we have $mN_s$ particles in total. We assign the number $k_{i} \equiv m(i-1)+k$ to the $k$-th particle of site $i$ ($i = 1,\dots,N_{s}; k=1,\dots,m$). The general Hamiltonian we
will consider is a linear combination of intra- and inter-site permutations:
\begin{align}
&H=\sum_i H_{(i)} + \sum_{i<j} H_{(i,j)},\\
\noalign{\noindent with}
&H_{(i)}= \sum_{k_i < l_i } J_{k_{i}l_{i}}P_{k_{i}l_{i}}\label{E: H_i}\\
\noalign{\noindent and}
&H_{(i,j)}= J_{ij}\sum_{k_i,l_j} P_{k_{i},l_{j}},
\end{align}
where $J_{k_{i}l_{i}}$ and $J_{ij}$ are the intra- and inter-site coupling constants, respectively. Note that the inter-site operator $H_{(i,j)}$ couples all the particles of site $i$ to all the particles
of site $j \neq i$ with the same coupling $J_{ij}$. Consequently, $H_{(i,j)}$ commutes with all intra-site permutations $P_{k_{i}l_{i}}$ and $P_{k_{j}l_{j}}$, and
of course with all $P_{k_{n}l_{n}}$ for $n\neq i,j$. It follows that $H_{(i,j)}$ commutes with all intra-site operators $H_{(i)}$. Moreover, the intra-site operators commute with each other. We thus obtain the following fundamental property:
\begin{eqnarray}
[H,H_{(i)}]=0, \quad \forall \enspace i=1,...,N_s. \label{E: commutation_H}
\end{eqnarray}
Therefore, we can diagonalize $H$ and all of the $H_{(i)}$ in a common basis. Accordingly, the Hilbert space can be partitioned into sectors corresponding to the set of
eigenvalues $e_i, (i=1,...,N_s)$ of the operators $H_{(i)}$, and the total Hamiltonian $H$ is block diagonal in this basis.

Since the Hamiltonians $H_{(i)}$ are linear combinations of permutations, they are $\mathrm{SU}(N)$ invariant. Hence, assuming that the spectrum of $H_{(i)}$ has no accidental degeneracy, each eigenvalue $e_i$ defines a local subspace
of the Hilbert space at site $i$ that belongs to a local irrep, which we will denote by $\beta(i)$. The Hamiltonian $H$ restricted to the corresponding block is the $\mathrm{SU}(N)$ model we want to study,
with irrep $\beta(i)$ at site $i$, and a reference energy $\sum_i e_i$.

To write the Hamiltonian in this block, we construct a projection operator $\mathrm{Proj}$ that maps onto the corresponding subspace. 
As we will show in subsection \ref{S: Proj}, the explicit form of such a projection operator can be derived exactly using general properties of the permutation group, without diagonalizing the local Hamiltonians $H_{(i)}$. Hence, our model can alternatively be characterized by the Hamiltonian $H'=\sum_{i<j} H_{(i,j)}$, restricted to the projected Hilbert space.

Note that the construction can easily be extended to the case where the number of particles $m_i$ at site $i$ varies from site to site. For simplicity in the examples to follow, we will focus on chains with the same number of particles at each site, i.e., $m_i=m$ independent of $i$.



\subsection{The $\mathrm{SU}(N)$ formulation of the problem}

In the most general case, an $\mathrm{SU}(N)$ Heisenberg-like interaction between two sites $i$ and $j$ can be written as:
\begin{equation} \label{E: Hij_gen}
H_{(i,j)}=J_{ij}\sum_{\mu,\nu} \hat{S}_{\mu\nu}^{i}\hat{S}_{\nu\mu}^{j},
\end{equation}
where the $\mathrm{SU}(N)$ generators satisfy at each site $i$ the commutation relation:
\begin{align} 
\label{commutation_relation}
\left[\hat{S}_{\alpha\beta}^{i},\hat{S}_{\mu\nu}^{i}\right] = \delta_{\mu\beta}\hat{S}_{\alpha\nu}^{i}-\delta_{\alpha\nu}\hat{S}_{\mu\beta}^{i}.
\end{align}

With $m$ particles per site, 
  the $\mathrm{SU}(N)$ generators for site $i$ are
\begin{equation} 
\label{SUN_generator_sitei}
\hat{S}_{\mu\nu}^{i}=\sum_{k_i=m(i-1)+1}^{mi}|\mu_{k_i}\rangle\langle \nu_{k_i}|-\frac{m\delta_{\mu,\nu}}{N}
\end{equation}
where the operator $|\mu_{k_i}\rangle\langle \nu_{k_i}|$ acts on the  particle $k_i$ to change its colour from $|\nu_{k_i}\rangle$ to $|\mu_{k_i}\rangle$.
The symbols $\mu_{k_i}$ and $\nu_{k_i}$ each stands for one of the $N$ colours, $A, B, C$, etc. 
The term $-\frac{m\delta_{\mu,\nu}}{N}$ renders the generators traceless.
This set of local generators  satisfies the commutation relation in Eq. \eqref{commutation_relation}.
The $\mathrm{SU}(N)$ Heisenberg interaction between two sites $i$ and $j$ shown in Eq. \eqref{E: Hij_gen} can then be rewritten as:
\begin{align}
\sum_{\mu,\nu} \hat{S}_{\mu\nu}^{i}\hat{S}_{\nu\mu}^{j}&=\sum_{\mu,\nu}\sum_{k_i,l_j}|\mu_{k_i}\rangle\langle \nu_{k_i}|\otimes|\nu_{l_j}\rangle\langle \mu_{l_j}|-\frac{m^2}{N} \nonumber \\
&=\sum_{k_i,l_j}\Big{\{}\sum_{\mu,\nu}|\mu_{k_i}\rangle \otimes |\nu_{l_j}\rangle \langle \nu_{k_i}|\otimes \langle \mu_{l_j}|\Big{\}}-\frac{m^2}{N}.
\end{align}
The constant $-m^2/N$ may be dropped.

The term inside braces is the permutation operator $P_{k_{i}l_{i}}$, which interchanges the $k$-th particle of site $i$ and the $l$-th particle of site $j$. Thus, the interaction Hamiltonian between sites $i$ and $j$ couples each of the $m$ particles at site $i$ to each of the $m$ particles at site $j$ [as shown in Fig. \ref{F: H_interact}\color{red}(a)\color{black} \, for the case of $m=3$ particles per site]:
\begin{equation}
 \label{E: Hij_perm}
H_{(i,j)} = J_{ij}\sum_{k_i,l_j} P_{k_{i},l_{j}},
\end{equation}
and the general $\mathrm{SU}(N)$ Heisenberg Hamiltonian is of the form
\begin{equation}
 \label{E: H_int}
  H_{\textup{interaction}} = \sum_{i<j} H_{(i,j)}. 
  \end{equation}

In particular, the Hamiltonian for an entire chain of $N_{s}$ sites, in which each site interacts with the site(s) adjacent to it, is simply
\begin{equation}
\label{E: H_int_2}
H_{\textup{interaction}} = \sum_{i = 1}^{N_{s} - 1} H_{(i,i+1)},
\end{equation}
in the case of open boundary conditions, where we have indexed adjacent sites with consecutive numbers, as demonstrated in Fig. \ref{F: H_interact}\color{red}{(b)}\color{black} \, for $N_{s} = 7$. We can treat periodic boundary conditions by adopting the computationally convenient indexing convention shown in Fig. \ref{F: H_interact}\color{red}(c) \color{black}, and writing
\begin{equation} \label{E: H_int_per}
H_{\textup{interaction}} = H_{(1,2)} + H_{(N_{s}-1,N_{s})} + \sum_{i=1}^{N_{s}-2}H_{(i,i+2)}.
\end{equation}

\begin{figure} \label{F: H_interact}
\centerline{\includegraphics[width=\linewidth]{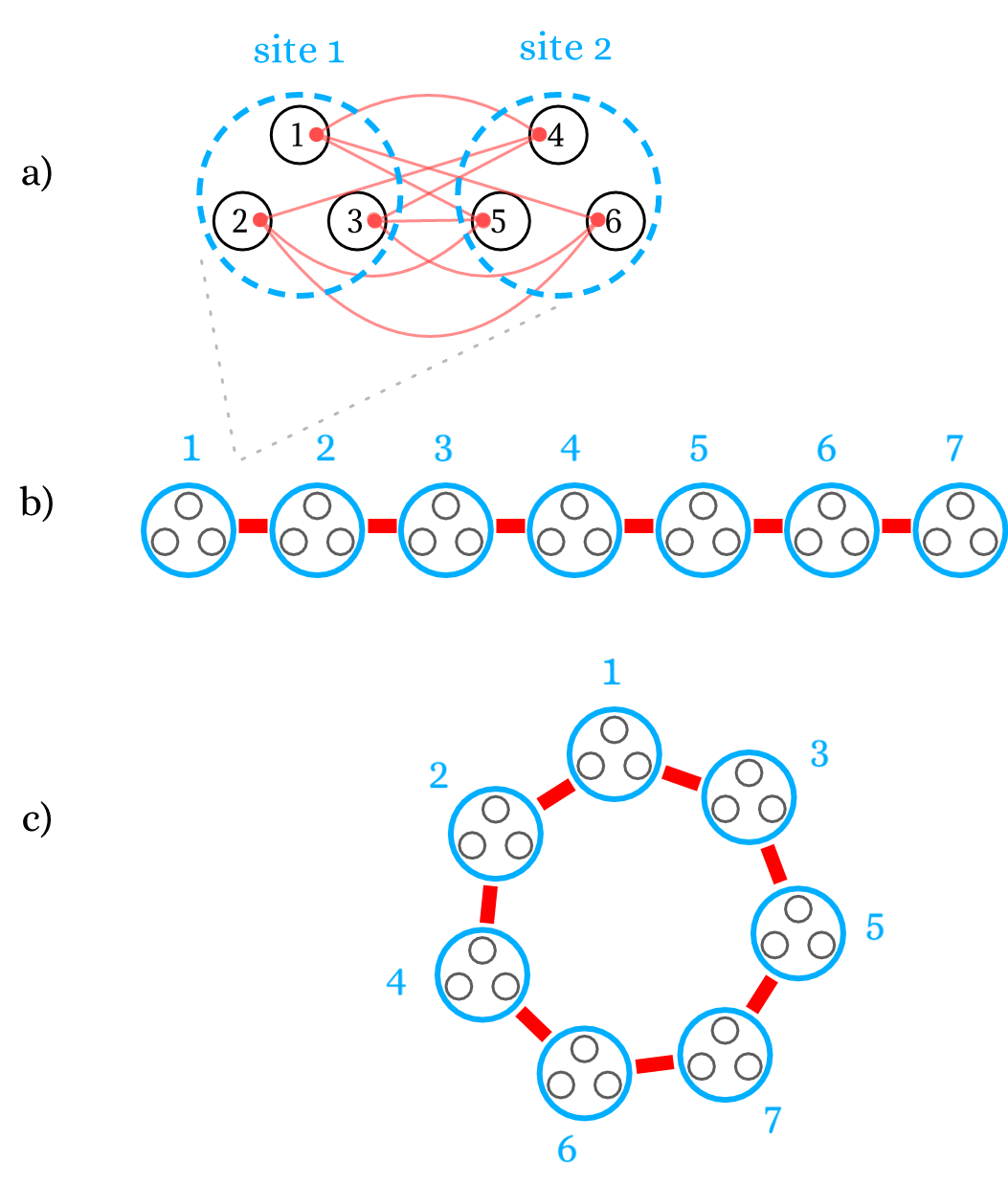}}
\caption{(a): The $\mathrm{SU}(N)$ Heisenberg interaction between sites 1 and 2 couples each of the particles of site 1, labelled 1, 2, and 3, to the particles of site 2, labelled 4, 5, and 6.
(b): Indexing scheme for sites on a chain with open boundary conditions. (c) Indexing scheme for periodic boundary conditions.}
\label{F: H_interact}
\end{figure}

\subsection{Projection operators} \label{S: Proj}
We now describe the procedure for writing the requisite projection operator as a linear superposition of permutations. Using a projection operator is essential: For a system of $N_{s}$ sites with $m$ particles per site, the full Hilbert space has dimension $N^{mN_{s}}$ and can be very large even for a small number of sites. Solving for the energies of an interaction Hamiltonian in this space would require diagonalizing a matrix of size $N^{mN_{s}} \times N^{mN_{s}}$, and the eigenvalues would include the spectra for \emph{all} possible combinations of local irreps.
Furthermore, this set of eigenenergies would encompass all of the different global irreps, since the full (reducible) Hilbert space can be decomposed as $\oplus_{\alpha}V^{\alpha}$, where the $\alpha$ are $\mathrm{SU}(N)$ irreps 
[refer to subection \ref{Appendix_irrep} for a review].
To obtain only the spectrum associated with a given global irrep $\alpha$ and a combination of specific local irreps, i.e., a given $\beta(i)$ at each site $i$, we must apply on the sector $V^{\alpha}$ the projection operator
\begin{equation} \label{E: Proj} \mathrm{Proj} = \prod_{i = 1}^{N_{s}}\mathrm{Proj}^{\beta(i)}(i). \end{equation}
$\mathrm{Proj}$ is formulated as a product of $N_{s}$ operators, one for each site, where $\mathrm{Proj}^{\beta(i)}(i)$ imposes the symmetry associated with $\beta(i)$ at site $i$.

Each $\mathrm{Proj}^{\beta(i)}(i)$ is a linear superposition of the $m!$ permutations among the $m$ particle numbers of site $i$. Its exact form can be determined analytically as follows.


The local eigenstates at site $i$ that belong to irrep $\beta(i)$ correspond to the eigenvectors of $\mathbf{H}_{(i)}^{\beta(i)}$, the matrix of the intra-site coupling Hamiltonian defined in Eq. \eqref{E: H_i}.
 We can write $\mathbf{H}_{(i)}^{\beta(i)}$ using the rules for the construction of Young's \textit{orthogonal representation} of the symmetric group, provided in appendix \ref{young_rep}. In this representation, the basis consists of the {\it orthogonal units} $o_{rs}^{\beta(i)}$ of $\beta(i)$. This basis of orthogonal units $o_{rs}^{\beta(i)}$ is directly related to the $f^{\beta(i)}$ standard Young tableaux (SYTs) of shape $\beta(i)$ [cf. Fig. \ref{F: Y_rules} in \ref{young_rep}]. 
It suffices for $\mathrm{Proj}^{\beta(i)}(i)$ to project onto just one of the eigenstates of $\beta(i)$. Thus, if we denote the orthonormal set of eigenvectors of $\mathbf{H}_{(i)}^{\beta(i)}$ by $\{\mathbf{v}_{1}^{\beta}, \dots, \mathbf{v}_{f^{\beta(i)}}^{\beta}\}$, the matrix representation for $\mathrm{Proj}^{\beta(i)}(i)$ can be very simply calculated as
\begin{equation} \label{E: vvT}
\mathbf{Proj}^{\beta(i)}(i) = \mathbf{v}_{j}^{\beta}\left(\mathbf{v}_{j}^{\beta}\right)^{T},
\end{equation} 
using any eigenvector $\mathbf{v}_{j}$ ($1 \leq j \leq f^{\beta(i)}$).

Then, to convert $\mathrm{Proj}^{\beta(i)}(i)$ from its matrix form above to a linear combination of permutation operators, we substitute the explicit form of the orthogonal units $o_{rs}^{\beta(i)}$ [given in appendix \ref{A: orthogonal_units}] into
\begin{equation} \label{E: Proj matrix elements} \mathrm{Proj}^{\beta(i)}(i) = \sum_{r,s} [\mathbf{Proj}^{\beta(i)}(i)]_{rs} o_{rs}^{\beta(i)}, \end{equation}
where $[\mathbf{Proj}^{\beta(i)}(i)]_{rs}$ is the matrix element in the $r$-th row and $s$-th column.

\bigskip

\begin{figure*}  \label{F: scheme}
\centerline{\includegraphics[width=\linewidth]{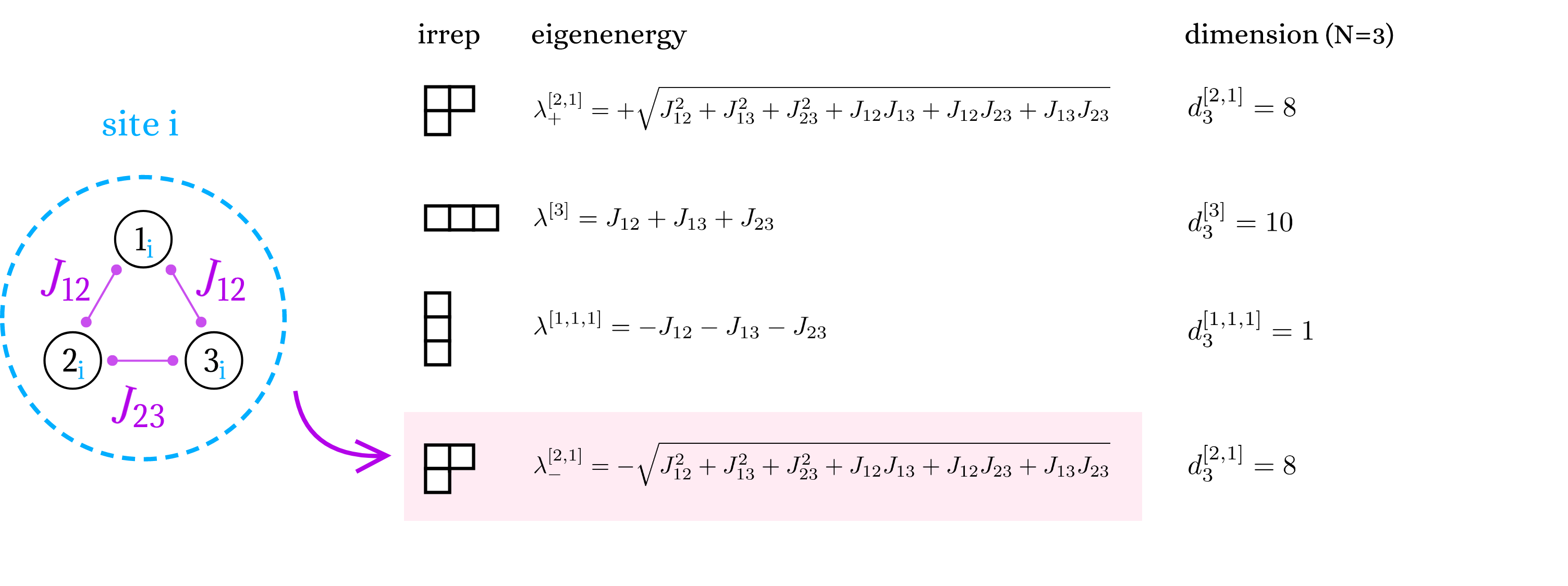}}
\caption{The irrep to which the ground state at site $i$ belongs can be determined from the spectrum of $H_{(i)}$, obtained for $m=3$ (and $N\geq 3$) by finding the eigenvalues of matrices \eqref{E: H3}-\eqref{E: H111}. For instance, the fundamental state is in the subspace associated with irrep $[2,1]$ if $\lambda_{-}^{[2,1]} < \min(\lambda^{[3]},\lambda^{[1,1,1]})$. In the case (shown) where $J_{12} + J_{13} + J_{23} > 0$, this occurs if the three coupling constants are such that $J_{12} < - J_{13}$, or $J_{12} > -J_{13}$ and $J_{23} < -\frac{J_{12}J_{13}}{J_{12}+J_{13}}$.}
\end{figure*}

In the example of $m=3$, the local Hamiltonian of site $i$ is \[H_{(i)} = J_{1_{i}2_{i}}P_{1_{i}2_{i}} + J_{1_{i}3_{i}}P_{1_{i}3_{i}} + J_{2_{i}3_{i}}P_{2_{i}3_{i}},\] and by applying the rules in subsection \ref{young_rep} to each of the three irreps $[3] =$ \ydiagram{3}, $[2,1] =$ \ydiagram{2,1}, and $[1,1,1] =$ \ydiagram{1,1,1} \enspace ($N \geq 3$), the matrix representations of $H_{(i)} $ are:

\begin{alignat}{2}
&\mathbf{H}_{(i)}^{[3]} &&= \begin{pmatrix} J_{12} + J_{13} + J_{23} \end{pmatrix} \label{E: H3} \\
&\mathbf{H}_{(i)}^{[2,1]} &&= 
	\begin{pmatrix}
		J_{12} - \frac{1}{2}J_{13} - \frac{1}{2}J_{23} 
		& -\frac{\sqrt{3}}{2}(J_{13} - J_{23}) \\
		-\frac{\sqrt{3}}{2}(J_{13} - J_{23}) 
		& -J_{12} + \frac{1}{2}J_{13} + \frac{1}{2}J_{23}
	\end{pmatrix} \label{E: H21} \\
&\mathbf{H}_{(i)}^{[1,1,1]} &&= \begin{pmatrix} -J_{12} - J_{13} - J_{23} \end{pmatrix} \label{E: H111}
\end{alignat}
where, for the sake of brevity, we have omitted the subscripts $i$ in the coupling constants. 
In particular, we can address the following question: given a set of constants $J_{k_{i}l_{i}}$, to which irrep does the ground state at site $i$ belong? This entails determining the irrep $\beta(i)$ for which the matrix $\mathbf{H}_{(i)}^{\beta(i)}$ has the lowest eigenvalue. Figure \ref{F: scheme} depicts an algebraic example.

Applying the above procedure [Eqs. \eqref{E: vvT} - \eqref{E: Proj matrix elements}] to the local irrep $\beta(i) = [2,1]$,  we find:
\begin{align} 
	\textup{Proj}_{\pm}^{[2,1]}(i) = &\pm \frac{1}{6\lambda^{[2,1]}}
		[(-2J_{12} + J_{13} + J_{23})P_{1_{i}2_{i}} \nonumber \\
		&+ (J_{12} - 2J_{13} + J_{23})P_{1_{i}3_{i}} \nonumber \\
		&+ (J_{12} + J_{13} - 2J_{23})P_{2_{i}3_{i}}] \nonumber \\
 		&+\frac{1}{6}(2Id - P_{1_{i}2_{i}}P_{2_{i}3_{i}} - P_{2_{i}3_{i}}P_{1_{i}2_{i}}), \label{E: Proj21}
\end{align}
where $P_{k_{i}l_{i}}$ is the permutation operator between particles $k_{i}$ and $l_{i}$, $Id$ is the identity operator, and $\lambda^{[2,1]} = \sqrt{{J_{12}}^{2} + {J_{13}}^{2} + {J_{23}}^{2} + J_{12}J_{13} + J_{12}J_{23} + J_{13}J_{23}}$.

This formula contains two projection operators, one obtained from each of the two eigenvectors of $\mathbf{H}_{(i)}^{[2,1]}$ via Eq. \eqref{E: vvT}: the positive sign corresponds to $\mathbf{v}_{+}^{[2,1]}(\mathbf{v}_{+}^{[2,1]})^{T}$, where $\mathbf{v}_{+}^{[2,1]}$ is the eigenvector with the positive eigenvalue $+\lambda^{[2,1]}$, while the negative sign corresponds to $\mathbf{v}_{-}^{[2,1]}(\mathbf{v}_{-}^{[2,1]})^{T}$. In general, for a local irrep $\beta(i)$ at site $i$, there are $f^{\beta(i)}$ local projection operators for site $i$, one for each of the $f^{\beta(i)}$ eigenstates of $\mathbf{H}_{(i)}^{\beta(i)}$. In practice, however, when we consider a Hamiltonian for interactions between different sites [Eq. \eqref{E: H_int}], only one of these $f^{\beta(i)}$ operators is required for each site, and the specific choice is inconsequential. Since the Hamiltonian for the internal coupling between particles of the same site, $H_{\textup{internal}}=\sum_{i=1}^{N_s}H_{(i)}$, commutes with the Hamiltonian $H = H_{\textup{internal}} + H_{\textup{interaction}}$  [cf. Eq. \ref{E: commutation_H}], the spectrum of $H$ is equal to the spectrum of $H_{\textup{interaction}}$ shifted by that of $H_{\textup{internal}}$. 

Therefore, if we have, for instance, the irrep $\beta = [2,1]$ at every site in a chain, the eigenvalues of $H_{\textup{internal}}$ can be found directly from the matrices $\mathbf{H}_{(i)}^{[2,1]}$ [Eq. \eqref{E: H21}], while to study $H_{\textup{interaction}}$, we can take either one of the two local operators in \eqref{E: Proj21}---say, $\mathrm{Proj}_{+}^{[2,1]}(i)$---and write $\mathrm{Proj} = \prod_{i = 1}^{N_{s}}\mathrm{Proj}_{+}^{[2,1]}(i)$ as our projection operator for the chain.

\bigskip

Furthermore, since $H_{\textup{interaction}}$ has no dependence on the local coupling constants $J_{k_{i}l_{i}}$, artificial values can be substituted for the coupling constants $J_{k_{i}l_{i}}$ into the \emph{general} formula for $\mathrm{Proj}^{\beta(i)}(i)$ (such as that in Eq. \eqref{E: Proj21} for $\beta(i)=[2,1]$) without affecting the final spectrum obtained for $H_{\textup{interaction}}$. In particular, if $J_{12} = 1$ and $J_{13} = J_{23} = 0$, the operators in Eq. \eqref{E: Proj21} reduce to the same forms as the orthogonal units $o_{11}^{[2,1]}$ and $o_{22}^{[2,1]}$ [cf. appendix \ref{A: orthogonal_units} for an explicit expression]. This result is general: given any local irrep $\beta(i)$ at site $i$, if we choose $J_{12}=1$ as the only non-vanishing constant, the $f^{\beta(i)}$ projection operators are equivalent to the explicit form of the orthogonal units $o_{rr}^{\beta(i)} (\forall r = 1,\dots, f^{\beta(i)})$, with the indices of the permutation operators adjusted to match the particle numbers of the site (i.e., $k \to k_{i} = m(i - 1) + k$). Any one of these operators can be used in the subsequent calculations for $H_{\textup{interaction}}$. In our implementation, for instance, we simply use $o_{11}^{\beta(i)}$ at every site $i$.

\bigskip

\section{The Algorithm} \label{S: algorithm}
\subsection{Equivalence classes of Young tableaux}
By construction, $\mathrm{Proj}$ is a product of $N_{s}$ operators $\mathrm{Proj}(i)$ ($i = 1,\dots, N_{s}$), each of which acts on one of the sites in isolation, i.e., each $\mathrm{Proj}(i)$ is composed of permutation operators that permute only the $m$ particles at site $i$. For example, if $\beta(1) = [2,1] = \beta(2)$, using $\mathrm{Proj}^{[2,1]}(i)=o_{11}^{[2,1]}$ for both sites, we obtain, for site 1,
\[ \mathrm{Proj}^{[2,1]}(1) = \frac{1}{6}(2Id + 2P_{12} -P_{13} - P_{23} - P_{12}P_{13} - P_{13}P_{12}) \] and, for site 2, 
\[\mathrm{Proj}^{[2,1]}(2) = \frac{1}{6}(2Id + 2P_{45} - P_{46} - P_{56} - P_{45}P_{46} - P_{46}P_{45}).\] The product of these two operators does not ever interchange a particle of site $1$ with a particle of site $2$.

Accordingly, we start by partitioning the $f^{\alpha}$ SYTs of shape $\alpha$ into \textit{equivalence classes}: Two SYTs are said to belong to the same equivalence class if for each of the $N_{s}$ sites, the $m$ numbers labelling the particles of that site occupy the same $m$ locations on both tableaux. In other words, for each equivalence class, the locations of the blocks for each site are fixed, and SYTs of that class differ only by rearrangements of the particle numbers of the same site within those fixed locations.

For instance, for $\alpha = [4,3,2]$ and $m = 3$ particles per site, all of the SYTs belonging to one of the equivalence classes are listed below. For each site $i$ ($i = 1, \dots, 3$), the three particle numbers $3(i-1)+1$, $3(i-1)+2$, and $3(i-1)+3$ occupy the same locations on all of the SYTs.
\begin{alignat*}{4}
& \ytableaushort{1234,568,79}*[*(lightblue)]{3}*[*(lightpink)]{3+1,2}*[*(lightgrey)]{0,2+1,2} \qquad && \ytableaushort{1234,568,79}*[*(lightblue)]{3}*[*(lightpink)]{3+1,2}*[*(lightgrey)]{0,2+1,2} \qquad && \ytableaushort{1234,568,79}*[*(lightblue)]{3}*[*(lightpink)]{3+1,2}*[*(lightgrey)]{0,2+1,2} \qquad && \ytableaushort{1234,568,79}*[*(lightblue)]{3}*[*(lightpink)]{3+1,2}*[*(lightgrey)]{0,2+1,2} \qquad \\[0.3cm]
&\ytableaushort{1235,468,79}*[*(lightblue)]{3}*[*(lightpink)]{3+1,2}*[*(lightgrey)]{0,2+1,2} \qquad
&&\ytableaushort{1235,469,78}*[*(lightblue)]{3}*[*(lightpink)]{3+1,2}*[*(lightgrey)]{0,2+1,2} \qquad
&&\ytableaushort{1236,457,89}*[*(lightblue)]{3}*[*(lightpink)]{3+1,2}*[*(lightgrey)]{0,2+1,2} \qquad
&&\ytableaushort{1236,458,79}*[*(lightblue)]{3}*[*(lightpink)]{3+1,2}*[*(lightgrey)]{0,2+1,2} \\[0.3cm]
&\ytableaushort{1236,459,78}*[*(lightblue)]{3}*[*(lightpink)]{3+1,2}*[*(lightgrey)]{0,2+1,2} && && &&\\[0.3cm]
\intertext{In total, there are 12 equivalence classes of $\alpha = [4,3,2]$ for $m = 3$. We can illustrate them schematically, using a different colour to indicate the three fixed locations for the particle numbers of each site, as below.}
\text{a) } &\ydiagram[*(lightblue)]{3}*[*(lightpink)]{3+1,2}*[*(lightgrey)]{0,2+1,2} \qquad
\text{b) } &&\ydiagram[*(lightblue)]{3}*[*(lightpink)]{3+1,1,1}*[*(lightgrey)]{0,1+2,1+1} \qquad
\text{c) } &&\ydiagram[*(lightblue)]{3}*[*(lightpink)]{0,3}*[*(lightgrey)]{3+1,0,2} \qquad
\text{d) } &&\ydiagram[*(lightblue)]{3}*[*(lightpink)]{0,2, 1}*[*(lightgrey)]{3+1,2+1,1+1} \\[0.3cm]
\text{e) } &\ydiagram[*(lightblue)]{2,1}*[*(lightpink)]{2+2,1+1}*[*(lightgrey)]{0,2+1,2} \qquad
\text{f) } &&\ydiagram[*(lightblue)]{2,1}*[*(lightpink)]{2+2,0,1}*[*(lightgrey)]{0,1+2,1+1} \qquad
\text{g) } &&\ydiagram[*(lightblue)]{2,1}*[*(lightpink)]{2+1,1+2}*[*(lightgrey)]{3+1,0,2} \qquad
\text{h) } &&\ydiagram[*(lightblue)]{2,1}*[*(lightpink)]{2+1,1+1,1}*[*(lightgrey)]{3+1,2+1,1+1} \\[0.3cm]
\text{i) } &\ydiagram[*(lightblue)]{2,1}*[*(lightpink)]{0,1+1,2}*[*(lightgrey)]{2+2,2+1} \qquad
\text{j) } &&\ydiagram[*(lightblue)]{1,1,1}*[*(lightpink)]{1+3}*[*(lightgrey)]{0,1+2,1+1} \qquad
\text{k) } &&\ydiagram[*(lightblue)]{1,1,1}*[*(lightpink)]{1+2,1+1,0}*[*(lightgrey)]{3+1,2+1,1+1} \qquad
\text{l) } &&\ydiagram[*(lightblue)]{1,1,1}*[*(lightpink)]{1+1,1+1,1+1}*[*(lightgrey)]{2+2,2+1}
\end{alignat*} 

\subsection{Basis states and matrix representation}
We will use $\mathcal{B}$ to denote the orthonormal basis spanning the projected Hilbert space we are looking for. Every basis state $\ket{\Psi} \in \mathcal{B}$ (sometimes called a {\it multiplet} \cite{weichselbaum2012}) satisfies:
\begin{equation} \label{E: basis} \mathrm{Proj} \ket{\Psi} = \ket{\Psi}. \end{equation}
Since the projection operator does not interchange particles belonging to different sites, it follows that each of these basis states can be identified with a superposition of SYTs that are in the same equivalence class. Indeed, if we generate all of the SYTs belonging to a certain class, we can directly extend Young's rules [cf. appendix \ref{young_rep}] to write $\mathrm{Proj}$ as a matrix in terms of this subset of SYTs. The basis state(s) belonging to this class can subsequently be found by solving for $\ket{\Psi}$ in Eq. \eqref{E: basis}. Repeating this for all of the equivalence classes with shape $\alpha$, we obtain a full basis for the particular subspace of $V^{\alpha}$ associated with the given combination of local irreps $\beta(i)$ $(i = 1,\dots,N_{s})$.

\bigskip

The number of basis states depends on the global irrep $\alpha$. Suppose we have local irrep $\beta(i)$ at site $i$ ($i = 1,\dots, N_{s})$. Let $D^{\alpha}(\mathcal{B})$ denote the number of states spanning basis $\mathcal{B}$ associated with a given $\alpha$. Then,
\[ \prod_{i = 1}^{N_{s}} d_{N}^{\beta(i)} = \sum_{\alpha} d_{N}^{\alpha} D^{\alpha}(\mathcal{B}). \]
Here, the sum $\sum_{\alpha}$ runs over all shapes $\alpha$ with $mN_{s}$ boxes and at most $N$ rows, and $d_{N}^{\alpha}$ and $d_{N}^{\beta(i)}$ are the dimensions of, respectively, the global irrep $\alpha$ and the local irrep $\beta(i)$ [cf. appendix \ref{Appendix_irrep}]. 

\bigskip

A major simplification can be made by noting that, given a specific combination of local irreps, only a certain subset of equivalence classes constitute basis states of the basis $\mathcal{B}$ for that combination of irreps. In the most general configuration with irrep $\beta(i)$ at site $i$, these "viable" classes are those that can be obtained when the Itzykson-Nauenberg\cite{itzykson} rules are applied to form the tensor product \[\bigotimes_{i = 1}^{N_{s}} \beta(i)\] in terms of Young diagrams. These rules can thus be implemented into an iterative scheme to obtain all of the viable classes for a given global irrep $\alpha$ and combination of local irreps $\beta(i)$ $(i = 1,\dots, N_{s})$. 

Furthermore, we can infer some general consequences of the rules that can be used to immediately identify invalid classes. For example, with irrep $\beta(i)$ at site $i$, a simple condition must always be satisfied: Let $r_{i}$ and $c_{i}$ denote the number of rows and number of columns, respectively, in the shape $\beta(i)$. Then, for any viable class, the $m$ blocks associated with the particles of site $i$ are situated on the tableaux such that there are no more than $c_{i}$ blocks in the same row, and no more than $r_{i}$ blocks in the same column, for all $i = 1, \dots, N_{s}$. The SYTs of classes for which this condition does not hold give zero upon projection onto the local irreps $\beta(i)$, and hence do not need to be considered. 

As an example, if each site is to be projected onto the irrep $[2,1]$, it follows from this condition that any class of tableaux on which the three blocks for some site are in the same row or in the same column are invalid. In the above diagram, then, tableaux a)-d), and j)-l) can all be neglected for this particular problem. By integrating this constraint with an optimized recursive algorithm, we are able to generate all 867893 of the viable equivalence classes of the global irrep $\alpha = [12,12,12]$ (i.e., the $\mathrm{SU}(3)$ singlet sector for 12 sites) with local irrep $\beta = [2,1]$ at each site in two minutes on single core of a standard CPU.

\bigskip

In the case of symmetric or anti-symmetric irreps at each site, there is only one basis state for each viable equivalence class, and it is possible to project a representative state of each class via a single, pre-determined formula \cite{PRBNataf2016}. For other symmetries, there may be more than one basis state associated with a given class, and the number of basis states may vary from class to class. In the general case, therefore, we simply solve Eq. \eqref{E: basis} for each class as a matrix equation (equivalent to finding the kernel of $(\mathrm{Proj} - \mathbf{I})\ket{\Psi}=\mathbf{0}$, where $\mathbf{I}$ and $\mathbf{0}$ are the identity matrix and the zero matrix).

Since the $\mathrm{Proj}$ does not permute particles between two different sites, the states $\ket{\Psi}$ in Eq. \eqref{E: basis} can be calculated more easily by finding the local states $\ket{\Psi_{i}}$ that satisfy
\begin{equation} \label{E: basis_i}
	\mathrm{Proj}^{\beta(i)}(i)\ket{\Psi_{i}} = \ket{\Psi_{i}}
\end{equation}
for each site $i$, then taking the tensor product \[\ket{\Psi} = \bigotimes_{i=1}^{N_{s}}\ket{\Psi_{i}}\] (and renormalizing). In solving equation \eqref{E: basis_i} for $\ket{\Psi_{i}}$, the largest matrix involved has dimension no greater than $m!$.

After finding the states of basis $\mathcal{B}$, it remains to write the interaction Hamiltonian with respect to this basis. 
This involves using the rules provided in appendix \ref{young_rep} to write the matrix coefficients of the permutation appearing in the interaction terms of Eq. \ref{E: Hij_perm}, then taking matrix-vector products to make the basis transformation.
In appendix \ref{S: find_H}, we give a provide a example and detail several useful technical simplifications that may be implemented to optimize the algorithm for this computation.

\section{Results} \label{S: results}
\begin{figure}
\centerline{\includegraphics[width=0.8\linewidth]{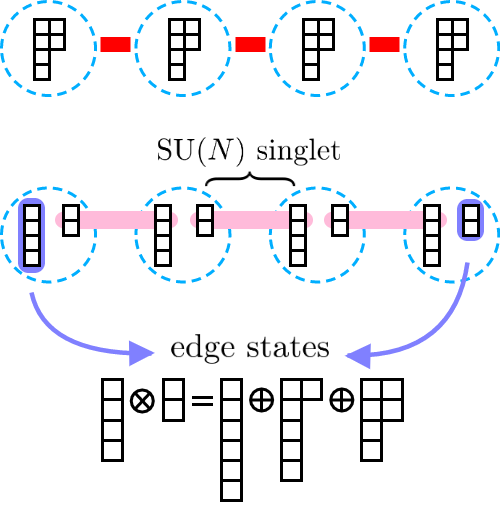}}
\caption{A chain of four sites with a two-column irrep on each site. The AKLT Hamiltonian is aimed at favouring the virtual decomposition into two one-column irreps and the recombination
over each link onto an $\mathrm{SU}(N)$ singlet. If the chain is open, the two edge irreps form edge states that live in the corresponding tensor product. \label{schema_AKLT}}
\end{figure}

We apply the method outlined above to investigate the presence of edge-states in $\mathrm{SU}(N)$ Heisenberg and AKLT Hamiltonians.
We first describe a procedure for building an $\mathrm{SU}(N)$ AKLT-like Hamiltonian, starting \textit{a priori} with the following irrep at each site:
\begin{equation}
\ytableausetup
{boxsize=0.75em}
N-p \left\{\raisebox{3.7 ex}{\begin{ytableau} $$ & $$ \\ $$ & $$ \\ $$  \\ $$ & \none[$\vdots$] \\ \none \\ \none[$\vdots$] \end{ytableau}}\right.   \raisebox{1.9ex}{$\Bigg\}p$},
\end{equation}
where $1\leq p\leq N/2$.
$p=1$ corresponds to the adjoint irrep, while $p=N/2$ represents the perfectly rectangular self-conjugate irrep.
At each site, the AKLT Hamiltonian 
favours 
the virtual decomposition of this local irrep into two fully antisymmetric irreps (i.e., two one-column irreps of lengths $N-p$ and $p$) and their recombination into a singlet over a link of the chain, as depicted in Fig. (\ref{schema_AKLT}). Consequently, we construct the AKLT Hamiltonian in such a way that it yields the minimal energy (arbitrarily set to zero) for any state whose local wave function over a link of the lattice lives in one of the $p+1$ irreps appearing in the tensor product of the two \textit{virtual} irreps (the one-column irreps of lengths $N-p$ and $p$). As an example, we consider such a tensor product for $N=6$ and $p=2$:
\begin{equation}
\label{tensor_virtual}
\ydiagram{1,1,1,1} \otimes \ydiagram{1,1} =\ydiagram{1,1,1,1,1,1} \oplus \ydiagram{2,1,1,1,1}\oplus  \ydiagram{2,2,1,1}.
\end{equation}
Moreover, the AKLT Hamiltonian should give a strictly positive energy for any irrep not present in the right-hand side of Eq. (\ref{tensor_virtual}) but present in the tensor product of two interacting sites: 
\begin{align}
\label{tensor_site}
\ydiagram{2,2,1,1} \otimes \ydiagram{2,2,1,1} &=\ydiagram{2,2,2,2,2,2} \oplus 2 \, \ydiagram{3,2,2,2,2,1}\oplus 3\, \ydiagram{3,3,2,2,1,1} \oplus
\ydiagram{3,3,2,2,2} \oplus ... 
\end{align}
For instance, in the latter decomposition, the Hamiltonian should give strictly positive energy for the irrep $[3,3,2,2,2]$ as well as for all other irreps not listed, but it should
give 0 for the first three irreps which, upon removing the first irrelevant column of length $N=6$, correspond to the right hand side of Eq. (\ref{tensor_virtual}).

To fulfill this set of requirements, one possibility is to take $H_{\text{AKLT}^N_p}= \sum_{i = 1}^{N_{s} - 1} H^{\text{AKLT}^N_p}_{(i,i+1)}$
where the link $(i,i+1)$ Hamiltonian $ H^{\text{AKLT}^N_p}_{(i,i+1)}$ is a product of $p+1$ terms, one per irrep (denoted $\beta_j$ below) appearing in the tensor product of the \textit{virtual} irreps [cf. Eq. \ref{tensor_virtual}]:
\begin{align}
\label{AKLT}
H^{\text{AKLT}^N_p}_{(i,i+1)}=\Pi_{\beta_j=\beta_1}^{\beta_{p+1}} \big\{H^{\text{tot}}(i,i+1)-C^2(\beta_j)\big\}.
\end{align}
$H^{\text{tot}}(i,i+1)$ is the sum of all the transpositions between any pairs of particles living in the ensembles of two sites $i$ and $i+1$: 
\begin{align}
H^{\text{tot}}(i,i+1)&=\sum_{ m(i-1)+1 \leq k<l \leq m(i+1) } P_{k,l}\nonumber
\\&=H_{(i,i+1)}^{J\equiv1}+H^{J\equiv1}_{(i)}+H^{J\equiv1}_{(i+1)}.
\end{align}
In other words, it is the sum of Hamiltonian appearing in Eq. (\ref{E: H_int_2}) and of the local atomic Hamiltonian for sites $i$ and $i+1$  of Eq. (\ref{E: H_i}) with all coupling constants $J_{k_i,l_i}$ and $J_{i,i+1}$ set to 1. 
$C^2$ is the quadratic Casimir of the $\mathrm{SU}(N)$ irrep.


Usually, a quadratic Casimir for $\mathrm{SU}(N)$ is written as a sum of products of generators of $\mathrm{SU}(N)$, but this approach is inconvenient here and we favour a permutation-like form of the quadratic Casimir, provided in the following equation.
For a general irrep $\alpha=[\alpha_1,\alpha_2,...,\alpha_k]$ with $n$ boxes, the quadratic Casimir can be calculated from the shape $\alpha$ as\begin{align}
\label{formula_casimir}
C^2(\alpha)=\sum_{1 \leq i<j \leq n} P_{i,j} =\frac{1}{2}\big\{ \sum_{i}\alpha_i^2-\sum_{j}(\alpha^T_j)^2\big\}
\end{align}
where the $\alpha_i$ are the lengths of the rows and the $\alpha^T_j$ are the lengths of the columns (which are also the rows of the transposed shape $\alpha^T$).
To apply the last formula in Eq. (\ref{AKLT}), it is important not to forget to add one column of $N$ boxes to the shapes $\beta_j$ appearing in the tensor product in Eq. (\ref{tensor_virtual}) in order for them to have exactly $m^2=N^2$ boxes in total.
So, for the example with $N=6$ and $p=2$, we obtain:
\begin{align}
\label{AKLT62}
H^{\text{AKLT}^6_2}_{(i,i+1)}= \big\{H_{(i,i+1)}+14\big\} \big\{H_{(i,i+1)}+8\big\} \big\{H_{(i,i+1)}+4\big\}, 
\end{align}
where we have set $H^{J\equiv1}_{(i)}=H^{J\equiv1}_{(i+1)}=C^2([2 2 1 1])=-5$ since we have the irrep $[2,2,1,1]$ at each site.
As a final step, one should check that the quadratic Casimir of all the other irreps (the ones appearing in Eq. (\ref{tensor_site}) but not in Eq. (\ref{tensor_virtual}))  are strictly larger than the ones appearing in Eq. (\ref{tensor_virtual}). For the example treated above, it is true.
In cases where it is not, to ensure the strict positivity of the last Hamiltonian on the states which would live locally (i.e., along a link) in one of the "bad" irreps, one could simply square the Hamiltonian (\ref{AKLT}) (or just part of it).

We have applied this logic to a simpler  case that has already attracted some attention in literature, 
for which $N = 4$ and $p = 2$. 
Then, since:
\begin{equation}
\label{tensor_virtual_2}
\ydiagram{1,1} \otimes \ydiagram{1,1} =\ydiagram{2,2} \oplus \ydiagram{2,1,1}\oplus  \ydiagram{1,1,1,1},
\end{equation}
and
\begin{align}
\label{tensor_site_2}
\ydiagram{2,2} \otimes \ydiagram{2,2} &=\ydiagram{2,2,2,2} \oplus  \ydiagram{3,2,2,1}\oplus \ydiagram{3,3,1,1}\nonumber \\
& \oplus
\ydiagram{4,2,2}  \oplus\ydiagram{4,3,1} \oplus \ydiagram{4,4},
\end{align}
one can write:
\begin{align}
\label{AKLT_SU4}
H^{\text{AKLT}^{4}_2}_{(i,i+1)}= \frac{1}{56}\big\{H_{(i,i+1)}+8\big\} \big\{H_{(i,i+1)}+4\big\} \big\{H_{(i,i+1)}+2\big\},
\end{align}
where we have added the normalization constant $1/56$ so that the linear term has amplitude one, in the development of the last product.
A relevant question is whether the $\mathrm{SU}(4)$ Heisenberg Hamiltonian for this irrep (which we can denote $H^{\text{Heis}^{4}_2}$)
holds the same quantum phase as $H^{\text{AKLT}^{4}_2}$. In particular, are the edge states, which are expected to be the states of lowest energy for $H^{\text{AKLT}^{4}_2}$, also the lowest energy states for $H^{\text{Heis}^{4}_2}$?
To answer this question, we have exactly diagonalized the interpolating Hamiltonian
\begin{align} \label{E: lambda}
H_{\lambda}=\lambda H^{\text{Heis}^{4}_2}+(1-\lambda) H^{\text{AKLT}^{4}_2}
\end{align} 
in each relevant irrep for different chain lengths up to a chain of 10 sites, for which the full Hilbert has a dimension of the order of $10^{13}$. 
The lowest energies for 10 sites are shown in Fig. (\ref{energiesSU4}): 
\begin{figure*}
\centerline{\includegraphics[width=0.9\linewidth]{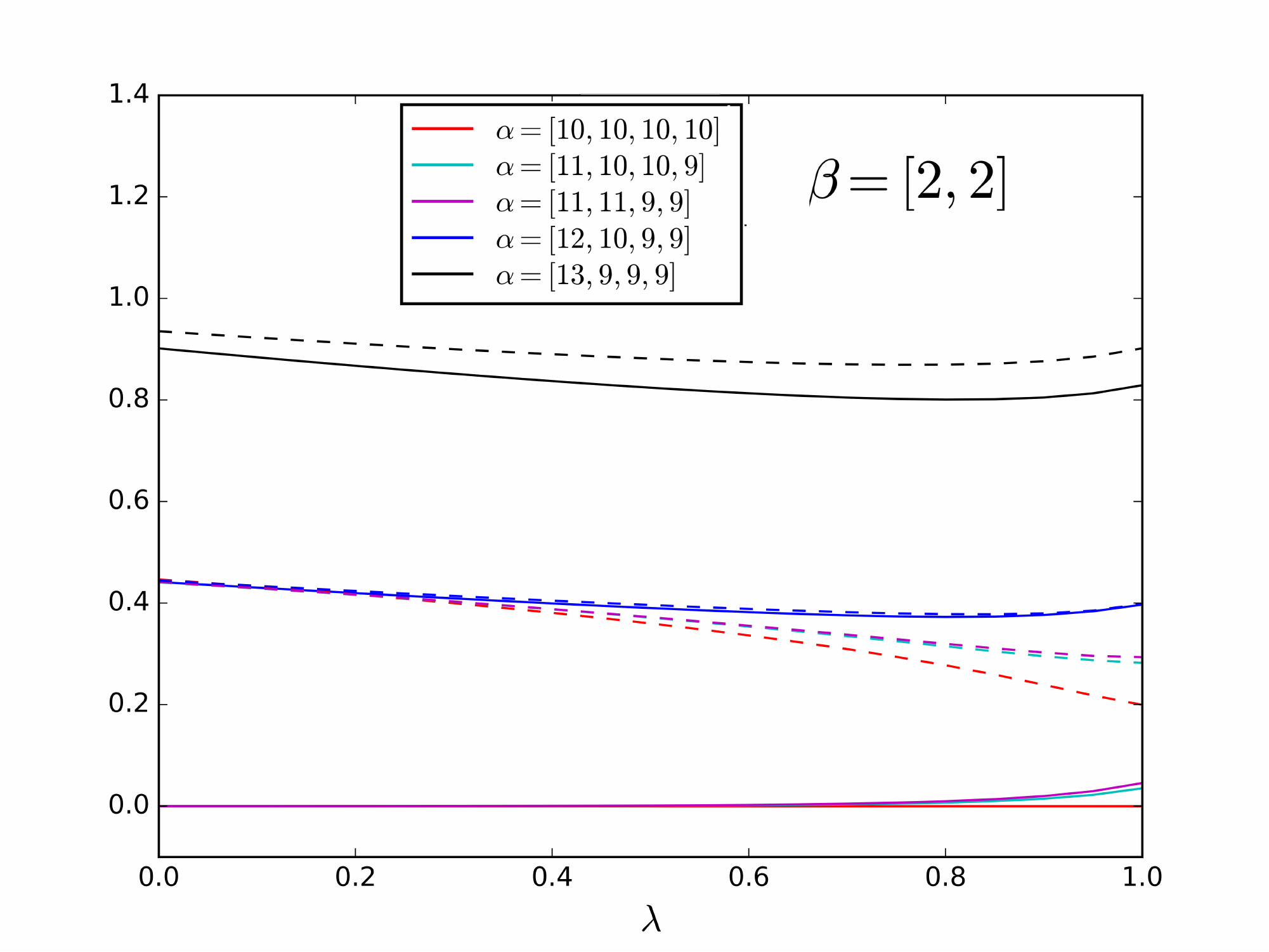}}
\caption{Lowest (solid) and second lowest (dashed) energies per site for each global irrep $\alpha$ with respect to the overall ground state energy (in $\alpha = [10,10,10,10]$) for an open chain of 10 sites with the irrep [2,2] at each site, as a function of $\lambda$. 
When $\lambda=0$, the Hamiltonian is purely AKLT; when $\lambda=1$, it is purely Heisenberg [cf. Eq. \eqref{E: lambda}]. The expected edge states live in the global irrep appearing in Eq. (\ref{tensor_virtual_2}), which, for a chain of $N_s$ sites, are the $\mathrm{SU}(4)$ singlet sector ($[N_s, N_s, N_s, N_s ]$), the adjoint ($[N_s+1, N_s, N_s, N_s-1 ]$) and the irrep equivalent to $[2,2]$, ($[N_s+1, N_s+1, N_s-1, N_s-1]$). It appears that the structure of the low energy states remains the same as we move from the AKLT to the Heisenberg point. \label{energiesSU4}} 
\end{figure*}
we see that the three lowest energy states are contained in the three irreps appearing in the RHS of Eq. (\ref{tensor_virtual_2}), and that they are well separated from the rest of the spectrum. This situation remains true as we tune $\lambda$ from the AKLT point ($\lambda=0$) to the Heisenberg point ($\lambda=1$), demonstrating continuity between the two points.

Interestingly, by calculating the energy spectrum in each global 
 irrep of the Hilbert space, our approach enables us to have a direct characterization of the edge states. In particular, the \textit{virtual decomposition} of each local irrep $[2,2]$ into two virtual $[1,1]$ irreps leads to the presence of edge states living in those latter irreps, the tensor product of which equal to the sum of the irreps contained in the ground state manifold, according to Eq. (\ref{tensor_virtual_2}).
These edge irreps $[1,1]$ form an effective and fractionalized representation of the symmetry group: mathematically, it corresponds to some \textit{projective} representation of the symmetry group. For $\mathrm{SU}(N)$, they can be classified according to the number of boxes (modulo $N$) of the virtual irreps (equal to $2$ in the example above), thus possibly giving rise to $N-1$ non-trivial topological phases\cite{Quella2013_phases}. Those phases (and the corresponding edge states), which have already received some attention in literature, mainly in the two aforementioned cases ($p=1$\cite{Katsura_2008,Orus_2011,Morimoto2014,Quella2015} and $p=N/2$\cite{Tanimoto2015} in the Hamiltonian (\ref{AKLT})) are usually characterized through other means 
 due to the difficulty of calculating the $\mathrm{SU}(N)$ quantum numbers when other numerical methods are used. 
For instance, one can discriminate those topological states by calculating some  non-local string order parameter\cite{quella2012,Quella2013_string,Morimoto2014,Tanimoto2015}, or the $Z_n$ Berry phase\cite{Motoyama_2015}.
Another method makes use of the entanglement spectrum\cite{Li_Haldane_2008} whose structure and degeneracies have been shown to reproduce that of the physical edge states \cite{Pollmann_2010}. In particular, in the case $N=4$ and $p=2$, the degeneracy of the edge irrep $[1,1]$ equal to $6$ has been observed in the entanglement spectrum of an Hamiltonian interpolating between an AKLT point and an Heisenberg point\cite{Tanimoto2015}, consistent with our results.
Our method reveals complementary and additional information.

\section{Conclusion and Perspectives} \label{S: conclusions}

In this paper, we have shown that the description of the basis states of $\mathrm{SU}(N)$ lattice models in terms of standard Young tableaux, which has previously been implemented
for the fundamental representation\cite{nataf2014} and for totally symmetric or antisymmetric representations\cite{PRBNataf2016}, can be extended to arbitrary irreducible
representations. The main difficulty as compared to the case of fully symmetric or antisymmetric irreps lies in the fact that the basis for each equivalence class of Young tableaux
is no longer of dimension 1, and special emphasis has been placed on the construction of this basis using projection operators defined in terms of elementary permutations.
As in previous cases, this approach allows one to work directly in irreps of the global $\mathrm{SU}(N)$ symmetry. Since for antiferromagnetic interactions the low lying states are typically
found in irreps whose dimensions are much, much smaller than that of the full Hilbert space, this enables us to reach sizes that are inaccessible to other formulations.

We have applied the method to the study of $\mathrm{SU}(4)$ Heisenberg chain with the irrep $[2,2]$ at each site, demonstrating that the ground state is of VBS type, and explicitly 
characterizing the irreps of the edge states, a very interesting by-product of this approach.

In the future, we aim to further develop our technique in three directions.
First, we hope to improve its efficiency in treating longer range interactions as well as three-site ring exchange interactions, in order to investigate 2D systems with exotic properties.
With one particle per site, $\mathrm{SU}(N)$ Heisenberg models on 2D systems have already been shown to possibly host chiral spin liquids with chiral edge states described by the $\mathrm{SU}(N)_1$ Wess-Zumino-Novikov-Witten conformal field theory \cite{natafPRL2016,natafrapidcomm2016,Dufour2016}.
To create non-Abelian Chiral Spin Liquids described by the $\mathrm{SU}(N)_k$ WZNW CFT (with $k>1$) requires more sophisticated irreps at each site of a 2D lattice \cite{hermele2009,Lecheminant2016,Lecheminant2016b}.
Secondly, we would like to generalize the use of SYTs to DMRG 
in order to take advantage of the $\mathrm{SU}(N)$ symmetry, an issue of considerable interest\cite{weichselbaum2012,mcculloch2007,mcculloch2002,gvidal2010} at present.
Finally, it would be interesting to determine whether it is possible to integrate both the complete $\mathrm{SU}(N)$ group and spatial symmetries in the same algorithm, as has been done for the special case of $N = 2$ \cite{schnack2009,schnack2010}.

\section{Acknowledgements}

We thank Akira Furusaki, Philippe Lecheminant, and Thomas Quella for useful discussions. This work has been supported by the Swiss National Science Foundation, and by the Research Internship program at EPFL. 

%
\section{Appendix}
\label{Appendix}
\subsection{Irreps of $\mathrm{SU}(N)$ }
\label{Appendix_irrep}

In general, for a system of $n$ particles, each irrep of $\mathrm{SU}(N)$ can be associated with a Young diagram composed of $n$ boxes arranged in at most $N$ rows. This represents a particular set of $\mathrm{SU}(N)$-symmetric $n$-particle wave functions. The shape $\alpha$ of the Young diagram is specified by a partition $\alpha = [\alpha_{1}, \alpha_{2}, \dots,\alpha_{k}]$ (with $1 \leq k \leq N$ and $\sum_{j=1}^{k} \alpha_{j} = n$), where the row lengths $\alpha_{j}$ satisfy $\alpha_{1} \geq \alpha_{2} \geq \dots \geq \alpha_{k} \geq 1$. The diagram can be filled with numbers 1 to $n$, and the resultant tableau is said to be \textit{standard} if the entries are increasing from left to right in every row and from top to bottom in every column. Standard Young tableaux (SYTs) play a central role in representation theory.

Using $[1]=\Box$ to denote the fundamental irrep, the set of all $n$-particle wave functions live in the full Hilbert space $\Box^{\otimes n}$. The multiplicity $f^{\alpha}$ of irrep $\alpha$ in this space is equal to number of SYTs with shape $\alpha$. This number can be calculated from the hook length formula,
\begin{equation}
f^{\alpha} = \frac{n!}{\prod_{i=1}^{n}l_{i}},
\end{equation}
where the hook length $l_{i}$ of the $i$-th box is defined as the number of boxes to the right of it in the same row, plus the number of boxes below it in the same column, plus one (for the box itself). The dimension $d_{N}^{\alpha}$ of the irrep can also be calculated from the shape as
\begin{equation} \label{E: dN}
 d_{N}^{\alpha} = \prod_{i=1}^{n}\frac{N+\gamma_{i}}{l_{i}},
\end{equation}
where $\gamma_{i}$ is the algebraic distance from the $i$-th box to the main diagonal, counted positively (resp. negatively) for each box above (resp. below) the diagonal. The full Hilbert space can be decomposed as 
\begin{equation} \label{E: sectors}
\Box^{\otimes n} = \oplus_{\alpha} V^{\alpha},
\end{equation} where $V^{\alpha}$ is the sector corresponding to irrep $\alpha$ (and if $d_{N}^{\alpha} > 1$, $V^{\alpha}$ can itself be decomposed into $d_{N}^{\alpha}$ equivalent subsectors, $V^{\alpha}= \oplus_{i=1}^{d^{\alpha}_N} V^{\alpha}_i $).
The equation for the dimension of the full Hilbert space thus reads
\begin{align}
\label{dimension_full}
N^{n}=\sum_{\alpha} f^{\alpha} d^{\alpha}_N,
\end{align}
where the sum runs over all Young diagrams with $n$ boxes and no more than $N$ rows.

As an example, for $n=2$, $\Box^2$ can be decomposed as a sum of the subspace spanned by the symmetric two-particle wave functions and that spanned by the antisymmetric two-particle wave functions: 
\[ \ytableausetup{smalltableaux}\ydiagram{1}^{\otimes 2} = \ydiagram{1} \otimes \ydiagram{1}= \ydiagram{2}  \oplus \ydiagram{1,1}. \]
There is only one SYT for each of the diagrams [2] and [1,1], so $f^{[2]}=f^{[1,1]}=1$, while Eq. \eqref{E: dN} gives $d_{N}^{[2]}=\frac{N(N+1)}{2}$ and $d_{N}^{[1,1]}=\frac{N(N-1)}{2}$. It is easy to check that Eq. (\ref{dimension_full}) is satisfied.

\subsection{Young's orthogonal representation of the symmetric group} \label{young_rep}
This subsection summarizes some useful results concerning the orthogonal representation of the symmetric group.

For a given tableau shape $\alpha$, a convenient representation of the symmetric group $\mathcal{S}_{n}$ can be formulated using Young's \textit{orthogonal units} $\{o_{rs}^{\alpha}\}_{r,s=1\dots f^{\alpha}}$. These are specific linear combinations of permutations, whose explicit forms are provided in section \ref{A: orthogonal_units} (for $n=3$ and $n=4$). They satisfy orthonormality:
\begin{equation}
o^{\alpha}_{rs}o^{\beta}_{uv} = \delta^{\alpha\beta}\delta_{su}o_{rv}^{\alpha} \quad \forall r,s = 1\dots f^{\alpha}, \forall u,v = 1\dots f^{\beta}
\label{orthonormal_relation} 
\end{equation}
as well as completeness:
\[ 
\sum_{\alpha} \sum_{r=1}^{f^{\alpha}} o^{\alpha}_{rr} = Id
\label{completeness_formula} 
\] 
and form a basis in which any linear superposition $\eta$ of permutations belonging to $\mathcal{S}_{n}$ can be uniquely decomposed as
\begin{equation}
\label{E: unique_decomposition}
 \eta = \sum_{\alpha,r,s} \mu_{rs}^{\alpha}(\eta)o_{rs}^{\alpha}, 
\end{equation}
where $\mu_{rs}^{\alpha}(\eta)$ are real coefficients.

An important result we will make frequent use of is that successive transpositions $P_{k,k+1}$, i.e., permutations between the consecutive numbers $k$ and $k+1$ ($1 \leq k \leq n-1$), takes an extremely simple form in the basis of orthogonal units. If we write $P_{k,k+1} = \sum_{\alpha,t,q} \mu_{tq}^{\alpha}(P_{k,k+1})o_{tq}^{\alpha}$, then, for a given shape $\beta$, the matrices $\bar{\bar{\mu}}^{\beta}(P_{k,k+1})$ defined by\[\left[\bar{\bar{\mu}}^{\beta}(P_{k,k+1})\right]_{tq} = \mu_{tq}^{\beta}(P_{k,k+1})\] are symmetric and orthogonal, and very sparse, with at most two nonzero entries in each row and in each column. These entries can be explicitly calculated as follows. We assign some fixed order (named {\it last letter order sequence}) to the $f^{\alpha}$ SYTs and label them $S_{1},\dots,S_{f^{\beta}}$. If $k$ and $k+1$ are in the same row (resp. column) on the tableau $S_{t}$, then $\mu_{tt}^{\beta}(P_{k,k+1}) = +1$ (resp. $-1$), and all other matrix elements involving $t$ vanish. If $k$ and $k+1$ are not in the same column nor the same row on $S_{t}$, and if $S_{q}$ is the tableau obtained from $S_{t}$ by interchanging $k$ and $k+1$, then the only non-vanishing matrix elements involving $t$ or $u$ are given by
\begin{equation} \label{E: Y_rules}
\begin{pmatrix} \mu_{tt}^{\beta}(P_{k,k+1}) \enspace & \mu_{tq}^{\beta}(P_{k,k+1}) \\[0.1cm] \mu_{qt}^{\beta}(P_{k,k+1}) \enspace & \mu_{qq}^{\beta}(P_{k,k+1}) \end{pmatrix} = \begin{pmatrix} -\rho \enspace & \sqrt{1-\rho^{2}} \\ \sqrt{1-\rho^{2}} \enspace & \rho \end{pmatrix}
\end{equation}
Here, $\rho$ is the inverse of the \textit{axial distance} from $k$ to $k+1$ on $S_{t}$, which is computed by counting $+1$ (resp. $-1$) for each step made downward or to the left (resp. upward or to the right) to reach $k+1$ from $k$. 

This simple yet incredibly useful formula is clarified in Figure \ref{F: Y_rules}. Moreover, since every permutation can be factorized into successive transpositions, we can use these rules to write the exact matrix representation of any permutation or linear superposition thereof via a few elementary calculations.

\begin{figure} \label{F: Y_rules}
\centerline{\includegraphics[width=0.7\linewidth]{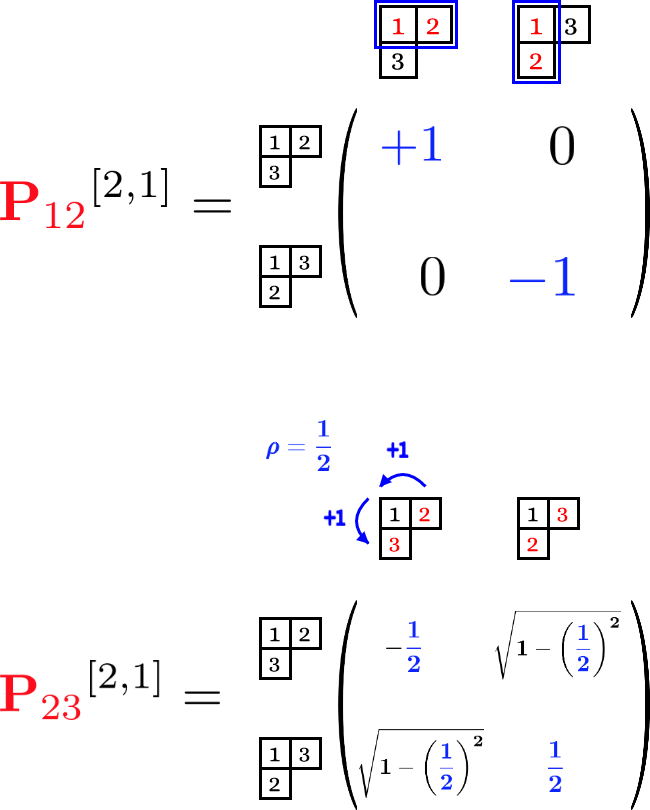}}
\caption{Writing the matrix representations of permutation operators $P_{12}$ and $P_{23}$ in the basis of SYTs of shape $\alpha = [2,1]$. We have labelled $\ytableaushort{12,3} = S_{1}$ and $\ytableaushort{13,2} = S_{2}$. For $P_{12}$, the numbers $1$ and $2$ are in the same row on $S_{1}$ and in the same column on $S_{2}$. For $P_{23}$, the axial distance between $1$ and $2$ on $S_{1}$ is 2, so $\rho = \frac{1}{2}$. $S_{2}$ is the tableau obtained from $S_{1}$ by interchanging $1$ and $2$, and we apply Eq. \eqref{E: Y_rules} accordingly. \label{F: Y_rules}}
\end{figure}

\subsection{Explicit form of the orthogonal units} \label{A: orthogonal_units}
We provide below the explicit expression of some orthogonal units for the irreps $[2,1]$ and $[2,2]$, as examples.

To construct the orthogonal units $
\left\{o^{\alpha}_{rs}\right\}, \forall r,s = 1\dots f^{\alpha},
$ for a given shape $\alpha$ with $n$ boxes, one can use the procedure of Thrall\cite{thrall}, which involves recursively constructing a product of antisymmetrizers and symmetrizers, adding one box at a time, for a total of $n$ boxes. 

We provide an alternative method that is useful when $n$ is not too large (typically, for $n\lesssim 10$).
For a given shape $\alpha$, we first generate all of the SYTs (using either the algorithm NEXYTB from Chap. 14 of Ref. \citen{nexytb} or the algorithm outlined in Appendix 5 of Ref. \citen{PRBNataf2016}). We then compute the $n!$ matrices of size $f^{\alpha}\times f^{\alpha}$ that represent the permutations $\eta$ of the symmetric group $\mathcal{S}_{n}$. The rules given in \ref{young_rep} can be used to write the $n-1$ matrices for transpositions between consecutive numbers, $P_{k,k+1}$ ($1\leq k\leq n-1$), and the remaining permutations of $\mathcal{S}_{n}$ can be obtained by decomposing them into transpositions and calculating the matrix products. Finally, we use the relation in Ref. \citen{rutherford}, which is the inverse of that shown in Eq. (\ref{E: unique_decomposition}):
\begin{align}
o^{\alpha}_{rs}=\frac{f^{\alpha}}{n!}\sum_{\eta \in \mathcal{S}_{n}} \mu^{\alpha}_{sr}(\eta^{-1}) \eta,
\end{align}
where $ \mu^{\alpha}_{sr}(\eta^{-1}) $ is the element in the $s^{th}$ row and $r^{th}$ column of the inverse of the matrix corresponding to permutation $\eta$.

The four orthogonal units associated for the shape [2,1] are thus linear combinations of the $3! = 6$ permutations of $\mathcal{S}_{3}$:
\begin{align} \label{E: orthogonal_units_21}
o^{[21]}_{11}&=\frac{1}{6}\left\{2Id+2P_{12}-P_{13}-P_{23}-P_{12}P_{13}-P_{13}P_{12}\right\} \nonumber \\
o^{[21]}_{12}&=\frac{1}{2\sqrt{3}}\left\{P_{23}-P_{13}-P_{12}P_{13}+P_{13}P_{12}\right\} \nonumber \\
o^{[21]}_{21}&=\frac{1}{2\sqrt{3}}\left\{P_{23}-P_{13}+P_{12}P_{13}-P_{13}P_{12}\right\} =\left(o^{[2,1]}_{12}\right)^{\dagger} \nonumber\\
o^{[21]}_{22}&=\frac{1}{6}\left\{2Id-2P_{12}+P_{13}+P_{23}-P_{12}P_{13}-P_{13}P_{12}\right\}.
\end{align}

while for the shape $[2,2]$, for instance, $o_{11}^{[2,1]}$ (which may directly be used as the projection operator onto the irrep associated with $[2,2]$) is
\begin{align*}
	o_{11}^{[2,2]} &= \frac{1}{24}(2Id+2P_{12}-P_{13}-P_{14}-P_{23}-P_{24}+2P_{34}\\
	&\quad -P_{12}P_{23}-P_{12}P_{24}+2P_{12}P_{34}+2P_{13}P_{24}-P_{13}P_{34}\\
	&\quad +2P_{14}P_{23}-P_{23}P_{12}-P_{23}P_{34}-P_{24}P_{12}-P_{34}P_{13}\\
	&\quad -P_{34}P_{23} +2P_{12}P_{13}P_{24}+2P_{12}P_{14}P_{23}-P_{12}P_{23}P_{34} \\
	&\quad -P_{12}P_{34}P_{23}-P_{23}P_{12}P_{34}-P_{34}P_{23}P_{12}).
\end{align*}

\subsection{Matrix representation of the interaction Hamiltonian}
 \label{S: find_H}
We discuss below several technical simplifications that may be implemented to optimize the construction of the matrix representing the permutation Hamiltonian.
We will focus on the specific example of the SU($3$) open chain of length $N_{s} = 5$ sites with local irrep $\beta = [2,1]$ at every site.
We will concentrate on the singlet sector (i.e of global irrep $[5,5,5]$).
 There are 16 viable equivalence classes, listed below. It will be found that each class may be spanned by one, two, or eight states, for a total of 32 basis states in $\mathcal{B}$.
\begin{alignat*}{4}
\text{a) } &\ydiagram[*(lightred)]{0,4+1,3+2}*[*(lightindigo)]{0, 3+1, 1+2}*[*(lightgrey)]{4+1, 2+1, 1}*[*(lightpink)]{2+2, 1+1}*[*(lightblue)]{2,1} \quad 
\text{b) } &\ydiagram[*(lightred)]{0,4+1,3+2}*[*(lightindigo)]{0, 2+2, 2+1}*[*(lightgrey)]{4+1, 0, 2}*[*(lightpink)]{2+2, 1+1}*[*(lightblue)]{2,1} \quad
\text{c) } &\ydiagram[*(lightred)]{0,4+1,3+2}*[*(lightindigo)]{4+1, 0, 1+2}*[*(lightgrey)]{0, 2+2, 1}*[*(lightpink)]{2+2, 1+1}*[*(lightblue)]{2,1} \quad
\text{d) } &\ydiagram[*(lightred)]{0,4+1,3+2}*[*(lightindigo)]{4+1, 3+1, 2+1}*[*(lightgrey)]{0, 2+1, 2}*[*(lightpink)]{2+2, 1+1}*[*(lightblue)]{2,1} \\[0.3cm]
\text{e) } &\ydiagram[*(lightred)]{0,4+1,3+2}*[*(lightindigo)]{0, 3+1,1+2}*[*(lightgrey)]{4+1, 1+2}*[*(lightpink)]{2+2,0,1}*[*(lightblue)]{2,1} \quad
\text{f) } &\ydiagram[*(lightred)]{0,4+1,3+2}*[*(lightindigo)]{0,2+2,2+1}*[*(lightgrey)]{4+1, 1+1,1+1}*[*(lightpink)]{2+2,0,1}*[*(lightblue)]{2,1} \quad
\text{g) } &\ydiagram[*(lightred)]{0,4+1,3+2}*[*(lightindigo)]{4+1,3+1,2+1}*[*(lightgrey)]{0, 1+2,1+1}*[*(lightpink)]{2+2,0,1}*[*(lightblue)]{2,1} \quad
\text{h) } &\ydiagram[*(lightred)]{0,4+1,3+2}*[*(lightindigo)]{0,3+1,1+2}*[*(lightgrey)]{3+2, 0,1}*[*(lightpink)]{2+1, 1+2}*[*(lightblue)]{2,1} \\[0.3cm]
\text{i) } &\ydiagram[*(lightred)]{0,4+1,3+2}*[*(lightindigo)]{4+1, 0, 1+2}*[*(lightgrey)]{3+1, 3+1,1}*[*(lightpink)]{2+1, 1+2}*[*(lightblue)]{2,1} \quad
\text{j) } &\ydiagram[*(lightred)]{0,4+1,3+2}*[*(lightindigo)]{4+1, 3+1, 2+1}*[*(lightgrey)]{3+1,0,2}*[*(lightpink)]{2+1, 1+2}*[*(lightblue)]{2,1} \quad
\text{k) } &\ydiagram[*(lightred)]{0,4+1,3+2}*[*(lightindigo)]{0, 3+1, 1+2}*[*(lightgrey)]{3+2, 2+1}*[*(lightpink)]{2+1, 1+1, 1}*[*(lightblue)]{2,1} \quad
\text{l) } &\ydiagram[*(lightred)]{0,4+1,3+2}*[*(lightindigo)]{0, 2+2, 2+1}*[*(lightgrey)]{3+2, 0, 1+1}*[*(lightpink)]{2+1,1+1,1}*[*(lightblue)]{2,1} \\[0.3cm]
\text{m) } &\ydiagram[*(lightred)]{0,4+1,3+2}*[*(lightindigo)]{4+1, 0, 1+2}*[*(lightgrey)]{3+1, 2+2}*[*(lightpink)]{2+1, 1+1, 1}*[*(lightblue)]{2,1} \quad 
\text{n) } &\ydiagram[*(lightred)]{0,4+1,3+2}*[*(lightindigo)]{4+1, 3+1, 2+1}*[*(lightgrey)]{3+1, 2+1, 1+1}*[*(lightpink)]{2+1, 1+1, 1}*[*(lightblue)]{2,1} \quad
\text{o) } &\ydiagram[*(lightred)]{0,4+1,3+2}*[*(lightindigo)]{3+2, 3+1}*[*(lightgrey)]{0, 2+1, 1+2}*[*(lightpink)]{2+1, 1+1, 1}*[*(lightblue)]{2,1} \quad
\text{p) } &\ydiagram[*(lightred)]{0,4+1,3+2}*[*(lightindigo)]{4+1, 3+1, 2+1}*[*(lightgrey)]{2+2, 2+1}*[*(lightpink)]{0, 1+1, 2}*[*(lightblue)]{2,1} 
\end{alignat*}

For instance, to write the the matrix of the Hamiltonian for the interaction between site 2 (particles 4, 5, and 6) and 3 (particles 7, 8, and 9),
\[ 
H_{(2,3)} = P_{47} + P_{48} + P_{49} + P_{57} + P_{58} + P_{59} + P_{67} + P_{68} + P_{69},
\]
[cf. Eq. \eqref{E: Hij_perm}], we consider, for each class, only the pink and grey blocks. 

For class a), the set of possible standard subtableaux for site 2 \enspace $\left( \ydiagram{1+2,1} \,\right)$ is \[S_{2} = \left\{\ytableaushort{\none 45, 6}\,,\, \ytableaushort{\none 46, 5}\,,\, \ytableaushort{\none 56, 4}\right\}. \] We write $\mathrm{Proj}^{[2,1]}(2) = \frac{1}{6}(2 Id + 2P_{45} - P_{46} - P_{56} - P_{45}P_{46} - P_{46}P_{45})$ with respect to these three subtableaux using Young's orthogonal representation, forming a $3 \times 3$ matrix $\mathbf{Proj}^{[2,1]}(2)$. 

Computing the kernel of $\mathbf{Proj}^{[2,1]}(2) - \mathbf{I}$ yields a single nonzero 3-dimensional vector: \begin{align*} \ket{\Psi^{a}_{2,1}} &\to \begin{pmatrix} u_{1} & u_{2} & u_{3} \end{pmatrix}^{T} \leftrightarrow u_{1} \ytableaushort{\none 45, 6} + u_{2} \ytableaushort{\none 46, 5} + u_{3} \ytableaushort{\none 56, 4}
\end{align*}

Performing a similar routine for site 3, with standard subtableaux $S_{3} = \left\{ \ytableaushort{\none\none\none\none7, \none\none 8, 9}\,,\, \ytableaushort{\none\none\none\none7, \none\none 9, 8}\,,\, \ytableaushort{\none\none\none\none8, \none\none 7, 9}\,,\, \right. $\\ $\left. \ytableaushort{\none\none\none\none8, \none\none 9, 7}\,,\, \ytableaushort{\none\none\none\none9, \none\none 7, 8}\,,\, \ytableaushort{\none\none\none\none9, \none\none 8, 7} \right\}$ yields two basis states represented as 6-dimensional vectors: 
\begin{align*}
\ket{\Psi^{a}_{3,1}} &\to \begin{pmatrix} v_{1} & \,v_{2} & \,\,v_{3} & \,\,v_{4} & \,v_{5} & \,v_{6}\end{pmatrix}^{T} \\
\ket{\Psi^{a}_{3,2}} &\to \begin{pmatrix}w_{1} & w_{2} & w_{3} & w_{4} & w_{5} & w_{6}\end{pmatrix}^{T}. 
\end{align*}

We then take the tensor products: Let \[\ket{\Psi^{a}_{2\otimes 3, 1}} = \ket{\Psi^{a}_{2,1}} \otimes \ket{\Psi^{a}_{3,1}} \to \begin{pmatrix} u_{1}v_{1} & u_{1}v_{2} & \dots u_{1}v_{6} & \dots & u_{3}v_{6}\end{pmatrix}^{T} \] and \[\ket{\Psi^{a}_{2\otimes 3, 2}} = \ket{\Psi^{a}_{2,1}} \otimes \ket{\Psi^{a}_{3,2}} \to \begin{pmatrix} u_{1}w_{1}, u_{1}w_{2}, \dots u_{1}w_{6}, \dots, u_{3}w_{6} \end{pmatrix}^{T}. \] These are the vectors we need to work with to write $H_{(2,3)}$ with respect to our basis. They each have 18 entries, corresponding to the set of 18 subtableaux formed by combining those of sites 2 and 3:
\begin{equation} \label{E: S2S3}
S_{2} \otimes S_{3} = \left\{\ytableaushort{\none\none457,\none68,9} \enspace, \dots, \ytableaushort{\none\none569,\none48,7}\right\}.
\end{equation}

In a similar fashion, the vectors representing the states $\ket{\Psi_{2\otimes 3}}$ can be found for the other classes. Each will be at most $6!^{2}=36$-dimensional. 

\bigskip

Then, the matrix elements of $\mathbf{H}_{(2,3)}$ in our 32-dimensional basis can be found the following way: We first write $H_{(2,3)}$ in the basis of subtableaux (please refer to the second bullet below) and then take the products 
\begin{equation} \label{E: matrix element} 
	\bra{\Psi^{x}_{2\otimes 3}}H_{(2,3)}\ket{\Psi^{y}_{2\otimes 3}},
\end{equation}
 where $x$ and $y$ run over all of the equivalence classes. 

The computational time and memory required can be reduced (by several orders of magnitude, for large systems) by taking advantage of the following key observations:
\begin{enumerate}
	\item Certain matrix elements are zero and need not be calculated. (In fact, the matrix representations for the $H_{(i,j)}$ in basis $\mathcal{B}$ are extremely sparse.) 
	
	Condition 1: In order for Eq. \eqref{E: matrix element} to be nonzero, the two classes $x$ and $y$ must be such that 6 blocks corresponding to sites 2 and 3 are in the same 6 locations on both classes. This follows from the fact that $H_{(2,3)}$ consists only of permutations between sites 2 and 3. Therefore, products between states of between class a) and those of class d) would automatically be zero, since the overall locations of blocks for sites 2 and 3 on class a), \ydiagram{2+3,1+2,1}, are different from those on class d), $\ydiagram{2+2,1+2,2}$. It can be seen that only classes g), j), n), and p) have the same total shape for sites 2 and 3 as class d). Additionally, between two classes that give the same subtableaux shape, there must be \textit{at most} one interchange between the three blocks of site 2 and those of site 3. For example, the subtableaux of class d), \ydiagram[*(lightgrey)]{0, 2+1, 2}*[*(lightpink)]{2+2, 1+1}, and of class g), \ydiagram[*(lightgrey)]{0, 1+2,1+1}*[*(lightpink)]{2+2,0,1}, differ only by an interchange of the leftmost pink block with the leftmost grey block. However, the subtableaux of class p), $\ydiagram[*(lightgrey)]{2+2,2+1}*[*(lightpink)]{0,1+1,2}$, differs from that of classes d) as well as g), j), and n) by \emph{two} interchanges. Hence, Eq. \eqref{E: matrix element} does not need to be calculated between the states of class p) and those of any other class.
	
	Condition 2: Moreover, classes $x$ and $y$ must be such that the locations of the blocks for each and every one of the \emph{other} sites are identical. To illustrate this more clearly, we consider a few classes of a slightly larger shape:
\begin{align*}
		\text{i)} \enspace & \ydiagram[*(lightpurple)]{5+2,5+1}*[*(lightred)]{0,4+1,3+2}*[*(lightindigo)]{0, 3+1, 1+2}*[*(lightgrey)]{4+1, 2+1, 1}*[*(lightpink)]{2+2, 1+1}*[*(lightblue)]{2,1} \qquad \text{ii)} \enspace \ydiagram[*(lightpurple)]{6+1,5+1, 4+1}*[*(lightred)]{5+1,4+1,3+1}*[*(lightindigo)]{0, 3+1, 1+2}*[*(lightgrey)]{4+1, 2+1, 1}*[*(lightpink)]{2+2, 1+1}*[*(lightblue)]{2,1} \\[0.3cm] \text{iii)} \enspace & \ydiagram[*(lightpurple)]{5+2,5+1}*[*(lightred)]{0,4+1,3+2}*[*(lightindigo)]{0, 3+1, 1+2}*[*(lightgrey)]{3+2, 2+1}*[*(lightpink)]{2+1, 1+1,1}*[*(lightblue)]{2,1}
\end{align*}
		Here, the blocks of sites 2 and 3 on all of above classes satisfy Condition 1. However, the blocks for site 5 and for site 6 are situated differently on class ii) than on classes i) and iii). As a result, products analagous to that in Equation \eqref{E: matrix element} between class ii) and class i) or between class ii) and iii) will all be zero.

	These two conditions follow from the nature of the interaction Hamiltonian and the orthogonality of the basis states in $\mathcal{B}$.
			
	\item Two vectors $\ket{\Psi^{x}_{2\otimes 3}}$, $\ket{\Psi^{y}_{2\otimes 3}}$ from different classes, i.e., $x \neq y$, are with respect to completely different bases. For class a), we found the set $S_{2} \otimes S_{3}$ [Eq. \eqref{E: S2S3}], which includes 18 subtableaux, but the analogue for class e) consists of 9 completely different subtableaux:
	\[ S^{e}_{2} \otimes S^{e}_{3} = \left\{\ytableausetup{smalltableaux}\ytableaushort{\none\none457,\none89,6}, \dots, \ytableaushort{\none\none569,\none78,4}\right\}. \]
	Then, to calculate $\bra{\Psi^{e}_{2\otimes 3,1}}H_{(2,3)}\ket{\Psi^{a}_{2\otimes 3,1}}$ and $\bra{\Psi^{e}_{2\otimes 3,1}}H_{(2,3)}\ket{\Psi^{a}_{2\otimes 3,2}}$ (there are two vectors for class a), as found above, and one for class e)), we first observe that $H_{(2,3)}$ is a sum of permutations between particles of sites 2 and 3, each of which can be decomposed into a product of transpositions with $P_{67}$ in the middle. For instance, $P_{69} = P_{79}P_{67}P_{79}$ and $P_{59} = P_{56}P_{79}P_{67}P_{79}P_{56}$. 
	
	Since matrix-vector products are linear, we take a single permutation in the sum at a time. Consider the product $\bra{\Psi^{e}_{2\otimes 3,1}}P_{59}\ket{\Psi^{a}_{2\otimes 3,2}} = \bra{\Psi^{e}_{2\otimes 3,1}}P_{56}P_{79}P_{67}P_{79}P_{56}\ket{\Psi^{a}_{2\otimes 3,2}}$. We find matrices for $P_{56}$ and $P_{79}$ with respect to each of the sets of subtableaux $S^{e}_{2} \otimes S^{e}_{3}$ and $S^{a}_{2} \otimes S^{a}_{3}$, by applying Young's rules. Then, instead of writing a full matrix representation for $P_{67}$, we place $S^{a}_{2} \otimes S^{a}_{3}$ on the columns and $S^{e}_{2} \otimes S^{e}_{3}$ on the rows (see below) to find a $9 \times 18$ matrix that can be used in the above product.
\begin{widetext}
\begin{equation*}
\mathbf{P}_{67} = \begin{blockarray}{cccccccc}
	\ytableaushort{\none\none457,\none68, 9} &\dots &\ytableaushort{\none\none458,\none69,7} &\ytableaushort{\none\none459,\none67,8} &\ytableaushort{\none\none459,\none68,7} &\dots &\ytableaushort{\none\none569,\none48,7} \\[0.3cm]
	\begin{block}{(ccccccc)c}
		0 & \dots & 0 & 0 & 0 & \dots & 0 & \enspace \ytableaushort{\none\none457,\none89,6} \\
  		0 & \dots & \sqrt{1-(\frac{1}{2})^{2}} & 0 & 0 & \dots & 0 & \enspace  \ytableaushort{\none\none458,\none79,6} \\
		0 & \dots & 0 & 0 & \sqrt{1 - (\frac{1}{2})^{2}} & \dots & 0 & \enspace \ytableaushort{\none\none459,\none78,6} \\
  		\vdots & \ddots & \vdots & \vdots & \vdots & \ddots & \vdots & \vdots \\[0.3cm]
  		0 & \dots & 0 & 0 & 0 & \dots & 0 & \enspace \ytableaushort{\none\none568,\none79,4} \\
  		0 & \dots & 0 & 0 & 0 & \dots& 0 & \enspace \ytableaushort{\none\none569,\none78,4}\\
	\end{block}
\end{blockarray}
\end{equation*}	
\end{widetext}
	
	\item For the interaction Hamiltonian between two given sites, we only need to work with the subtableaux composed of the $2m$ blocks of those two sites. Some of these sets of subtableaux will be the same for multiple classes. For example, among the 20 classes in our example, the blocks for site 1 and site 2 are the same on classes a) through d): \ydiagram[*(lightblue)]{2,1}*[*(lightpink)]{2+2,1+1}. Hence, when calculating $H_{(1,2)}$, it suffices to find the vectors and resultant matrix elements for such a configuration just once, instead of repeating the computation.
	
	\item Finding the kernel of $\mathbf{Proj}(i) - \mathbf{I}$ for each site $i$ individually and forming the tensor product(s) between the resultant vectors of two sites may not account for all of the basis states, since, as we have seen, some sites give rise to more than one vector. However, when writing the matrix for $H_{i,i+1}$, it suffices to find the vectors for the sites $i$ and $i+1$ and repeating their matrix elements in certain positions in the matrix (so that the same ordered basis is used for all the $H_{i,i+1}$). These positions depend on the number of vectors that correspond to the sites $j$ for $j < i$ and for $j > i + 1$.
	
\end{enumerate}

It should be noted that the list of simplifications provided above is vital, but by no means exhaustive.


\bibliographystyle{apsrev4-1}
\color{black}\bibliography{ED_general.bib}

\end{document}